\documentclass[draft]{agujournal2019}
\journalname{AGU}
\usepackage{url} %this package should fix any errors with URLs in refs.
\usepackage{lineno}
\usepackage[inline]{trackchanges} %for better track changes. finalnew option will compile document with changes incorporated.
\usepackage{soul}
\usepackage{amssymb}
\usepackage{amsmath}
\usepackage{mathtools}
\usepackage{float}

\begin{document}

%% ------------------------------------------------------------------------ %%
%  Title
%
% (A title should be specific, informative, and brief. Use
% abbreviations only if they are defined in the abstract. Titles that
% start with general keywords then specific terms are optimized in
% searches)
%
%% ------------------------------------------------------------------------ %%

% Example: \title{This is a test title}
\justify
\title{A Bayesian Deep Learning Approach to Near-Term Climate Prediction}

%% ------------------------------------------------------------------------ %%
%
%  AUTHORS AND AFFILIATIONS
%
%% ------------------------------------------------------------------------ %%

% Authors are individuals who have significantly contributed to the
% research and preparation of the article. Group authors are allowed, if
% each author in the group is separately identified in an appendix.)

% List authors by first name or initial followed by last name and
% separated by commas. Use \affil{} to number affiliations, and
% \thanks{} for author notes.
% Additional author notes should be indicated with \thanks{} (for
% example, for current addresses).

% Example: \authors{A. B. Author\affil{1}\thanks{Current address, Antartica}, B. C. Author\affil{2,3}, and D. E.
% Author\affil{3,4}\thanks{Also funded by Monsanto.}}

\authors{Xihaier Luo\affil{1}, Balasubramanya T. Nadiga\affil{2}, Yihui Ren\affil{1}, Ji Hwan Park\affil{1}, Wei Xu\affil{1}, and Shinjae Yoo\affil{1} }
\affiliation{1}{Computational Science Initiative, Brookhaven National Laboratory, Upton, NY 11973, USA} 
\affiliation{2}{Los Alamos National Laboratory, Los Alamos, NM 87545, USA}

%% Corresponding Author:
% Corresponding author mailing address and e-mail address:

% (include name and email addresses of the corresponding author.  More
% than one corresponding author is allowed in this LaTeX file and for
% publication; but only one corresponding author is allowed in our
% editorial system.)

% Example: \correspondingauthor{First and Last Name}{email@address.edu}

\correspondingauthor{Xihaier Luo}{xluo@bnl.gov}

%% Keypoints, final entry on title page.

%  List up to three key points (at least one is required)
%  Key Points summarize the main points and conclusions of the article
%  Each must be 140 characters or fewer with no special characters or punctuation and must be complete sentences

% Example:
% \begin{keypoints}
% \item	List up to three key points (at least one is required)
% \item	Key Points summarize the main points and conclusions of the article
% \item	Each must be 140 characters or fewer with no special characters or punctuation and must be complete sentences
% \end{keypoints}

\begin{keypoints}
\item Model bias and associated initialization shock are serious shortcomings that reduce prediction skill in state-of-the-art decadal climate prediction efforts.
 \item A complementary machine-learning-based approach to climate prediction is considered. Both deterministic and probabilistic machine learning approaches are examined.
\item In addition to providing useful measures of
  predictive uncertainty, Bayesian versions of deep learning models outperform
  their deterministic counterparts in terms of predictive skill.
\end{keypoints}

\begin{abstract}
  Since model bias and associated initialization shock are serious
  shortcomings that reduce prediction skill in state-of-the-art
  decadal climate prediction efforts, we pursue a complementary
  machine-learning-based approach to climate prediction. The example
  problem setting we consider consists of predicting natural
  variability of the North Atlantic sea surface temperature on the
  interannual timescale in the pre-industrial control simulation of
  the Community Earth System Model (CESM2). While previous works have
  considered the use of recurrent networks such as convolutional LSTMs
  and reservoir computing networks in this and other similar problem
  settings, we currently focus on the use of feedforward convolutional
  networks.  In particular, we find that a feedforward convolutional
  network with a Densenet architecture is
  able to outperform a convolutional LSTM in terms of predictive
  skill.  Next, we go on to consider a probabilistic formulation of
  the same network based on Stein variational gradient descent and find that
  in addition to providing useful measures of predictive uncertainty,
  the probabilistic (Bayesian) version improves on its deterministic
  counterpart in terms of predictive skill. Finally, we characterize
  the reliability of the ensemble of ML models obtained in the
  probilistic setting by using analysis tools
  developed in the context of ensemble numerical weather prediction.
\end{abstract}

\vspace{0.1cm}

\section*{Plain Language Summary}
Businesses and government agencies rely heavily on numerical predictions of climate variables such as temperature and precipitation for a wide variety of purposes ranging from integrated assessment to developing mitigation strategies to developing resilience and adaptation strategies. Developing interannual to decadal predictions using comprehensive and complex climate and earth system models, however, are computationally intensive. As such, computationally efficient and accurate surrogates of comprehensive earth system models is highly desired. Data-driven models using advanced deep learning algorithms are promising for this purpose. This paper first considers a recently proposed convolutional network architecture to develop such a surrogate and then integrates Bayesian inference to this architecture to further assess predictive uncertainty. We show that the resulting Bayesian deep learning model not only improves prediction accuracy but also quantifies the uncertainty arising from the data and model.

\newpage
\section{Introduction}
\label{sec1}
The climate system consists of diverse yet interconnected components, such as the atmosphere, oceans, etc., and each can exhibit complex, multiscale, and chaotic behaviors.
Additional interactions and feedbacks among these subsystems drive dynamic evolution over an enormous range of spatial and temporal scales in the climate system \cite{change2007climate, ipcc2013, ipcc21}.
In this setting, comprehensive climate models have emerged as a powerful tool in helping unravel and better comprehend the myriad processes underlying climate and climate change.
Moreover, studies using such models have greatly improved the understanding of climate system processes over the past few decades~\cite{ipcc21}.

Importantly, comprehensive climate models have helped to better anticipate the climate system's response to external forcings, such as those stemming from increased greenhouse gases that typically are realized on a timescale of a few decades or longer.
At shorter timescales at which natural variability plays an increasingly important role, however, improvements in the ability to predict climate are {\em not} commensurate with advances in understanding dynamics and processes  \cite{change2007climate, ipcc2013, ipcc21}.
Notably, improvements in predicting the El Niño-Southern Oscillation (ENSO) remain more of an exception than the rule.
Poor predictive skill at the shorter timescales is due to the fact that sources of predictability at these timescales reside in modes of natural variability of the climate system, and because models have difficulty in representing and capturing such modes and their timing with adequate accuracy.

Modes of natural variability in the climate system often are associated with delicate balances between multiple physical processes, and realizing those same balances in a climate model is difficult.
This leads to biases in a model's representation of the modes of variability.
These model biases also exist in the representation of the mean climate state.
While the downstream dynamical consequences of such model biases tend to be both complicated and manifold, from a dynamical systems perspective, an overall consequence tends to be that the model attractor is biased as well.

One way to understand poor predictive skill at shorter timescales is in terms of bias in the model's representation of the climate attractor:
when initialized predictions attempt to realize the predictability associated with natural variability by initializing the model state to be consistent with an observed climate state, the biased model attractor quickly pulls it away.
This leads to the model trajectory exhibiting a jump away from the observed trajectory toward the biased model attractor that typically involves complicated nonlinear dynamics.
An invariable effect tends to be loss of predictive skill \cite{nadiga2019enhancing}.

The current approach for dealing with this loss of skill consists of statistically correcting the predictions in a post-processing step.
Given the nonlinear and complicated dynamics that take place to effect the dramatic readjustment of the flow field (e.g., \cite{sanchez2016drift}), namely, the jump-like behavior of the initialized prediction trajectory, it is unlikely that the statistical post-processing of the predictions is capable of correctly compensating for these dynamics.

Given the problems associated with a comprehensive climate-model-based approach to near-term predictions, we are interested in investigating and developing alternative data-driven approaches to such predictions.
Herein, we split future climate into ``near-term'' and ``long-term'' and define near-term to mean the period over which initial conditions (IC) matter.
Thus, while long-term predictability is solely determined by boundary conditions (BC) and/or forcing, near-term predictability is affected by both BC/forcing and IC.

In particular, we are investigating the utility of an approach for
predicting near-term variations in a quantity of interest (QoI) that
is based on learning spatiotemporal variability of that QoI in a
controlled setting.  Such learning can be achieved using both
feedforward and recurrent neural networks (FNN, RNN respectively) (and transformer networks that are
beginning to outperform RNNs in at least certain
applications). Using RNNs for learning spatiotemporal
variability can be traced back to applying optical flow-based computer
vision techniques to extrapolate radar echo images toward {\em
  nowcast}ing convective precipitation (e.g., see
\cite{sakaino2012spatio}). Further developments along these lines
wherein precipitation nowcasting is formulated in the general
framework of a ``sequence-to-sequence'' learning
problem---transforming a sequence of past radar maps to a sequence of
future radar maps---quickly led to the proposal of a convolutional
long short-term memory (convLSTM) architecture/approach
\cite{xingjian2015convolutional}. In essence, a convLSTM network
determines the future states of a QoI at a spatial location using past
states of a local neighborhood and other inputs. Subsequently,
convLSTM has emerged as a machine learning (ML) technique that
delivers good performance in various applications.  This is especially
evident in some previous work involving climate-relevant settings of
predicting interannual variations of global surface temperature and
sea-surface temperature in ocean basins \cite{nadiga2019predicting,
  parkmachine, jiang2019interannual}. As such, even as other recurrent
neural network (RNN) architectures have emerged in the context of
sequence-to-sequence learning (e.g., attention-based transformers) and
are displacing convLSTM as the state-of-the-art, this work is
restricted to considering feedforward architectures and comparing
their performance to convLSTM. We will report on ongoing work using
attention-based methods elsewhere.

Another contribution of the present work consists of considering the
ML-based prediction of near-term climate variations in a probablistic
fashion as opposed to a deterministic approach.  In the context of
numerical weather prediction (NWP), the chaotic nature of atmospheric
dynamics necessitates considering the evolution of an ensemble of
trajectories in order to make reliable forecasts (of the one
trajectory that actually is realized in the observed weather system).
Starting with the statistical-dynamical prediction methods of
\cite{epstein1969stochastic} and more widely adopted at NWP centers
across the globe since the early 1990s\footnote{The European Center
  for Medium-Range Weather Forecasts (ECMWF) has been leading the
  charge.}, probabilistic forecasts using an ensemble prediction
system (EPS) have proven to be valuable in improving the skill of
weather forecasts.

Likewise, we expect that probablistic ML models of spatiotemporal
variability of climate will be both more skillful and useful than
deterministic ML models.  However, probabilistic ML remains in its
infancy. As such, developing and applying efficient probabilistic deep
learning models is difficult, and studies examining their utility and
performance are few.  In this context, assuming the network parameters
(weights and biases) are random variables and applying Bayes' rule
provide the theoretical basis for inferring the posterior distribution
of the network parameters that best fit the training data.  Here, we
note that (parametric) variational inference (VI) methods were
developed to efficiently approximate such inference computationally by
minimizing the Kullback-Leibler (KL) divergence between an approximate
posterior and the true posterior.  Subsequently, to extend the use of
VI beyond the specialized families of distributions that enjoy
particular conjugacy properties, approaches to nonparametric VI have
been developed.  Our study considers the Stein variational gradient
descent (SVGD) approach to nonparametric VI.  By adapting and applying
this probablistic deep learning approach to the climate prediction
problem being considered, we find that as in the context of NWP, a
probabilistic ML approach serves to improve on the skill of a
deterministic ML approach.

Next, we comment on the nature of the ensemble in a probabilistic ML
setting.  For this, it is useful to note that in the NWP setting---a
first-principles-based setting---two kinds of ensembles have 
typically been used in EPSs: (1) Initial condition ensembles (ICE) where
the model (is assumed to be perfect and so the model) configuration is
held fixed and uncertainty in estimation of the state of the system is
represented by an ensemble of initial conditions. (2) Perturbed
physics ensembles (PPE) where the initial condition is held fixed, but
the parameterizations that are used to represent unresolved processes
are perturbed to represent uncertainty related to model
error.
In the current data-driven probabilistic ML setting,
uncertainty represented by the ensemble may be thought of in the PPE
sense as the diversity of predictions can be traced back to
perturbations of the weights and biases that constitute the model's
{\em parameters}.  In this data-driven setting, while it is true that
the probabilistic learning algorithm is trying to learn generalities
over a diverse set of IC to infer the perturbations of the ML model
parameters that are required, the IC diversity in the training data is
{\em not} a representation of uncertainty in state estimation as would
be required for an ICE.

Finally, we make novel use of diagnostics developed for NWP-EPS in an
ML context. This is motivated by the fact that the goal of ensemble
prediction, whether it is in the more traditional context of ensemble
prediction systems or in the current probabilistic ML context, is for
the prediction to span the range of likely outcomes given the
uncertainties \cite{leith1974theoretical}. These diagnostics are based on the joint
analysis of error and ensemble variance.  To the best of our
knowledge, we use these diagnostics for the first time in the context
of probabilistic ML to gain added insight into both the network
architecture and the process of probabilistically inferring the
weights of the network. In this context, we note, however, that the
joint analysis of error and ensemble variance can be carried out in
many ways and we consider only the most elementary/simplest of such
methods.  Using the error-spread and rank-histogram diagnostics, we
find, in a global sense, that the ML prediction ensemble is
underdispersed. And the behavior persists on enlarging the size of the
ensemble. This leads us to further considering the reliability
diagnostics in a spatially localized or fine-grained sense. On so
doing, a more complicated picture emerges: There are some regions,
such as the subpolar North Atlantic, where the ML ensemble is actually
overdispersed. However there are other larger regions, such as
equatorial and tropical North Atlantic, where the ensemble is
underdispersed. Therefore, in the aggregate an overall underdispersive
behavior emerges. As such making changes to the probabilistic ML
methodology to further improve the reliability of the prediction
ensemble and making it optimally reliable tends to be tricky.

The rest of the paper is organized as follows. Section \ref{sec2} presents the details of the data and definitions of prediction problem. Section \ref{sec3} discusses the proposed Bayesian deep learning model, including the key derivation, architecture designs, training and testing procedures, and implementation guidelines. Section \ref{sec4} performs a comparative study on different models and covers a qualitative and quantitative examination of the climate predictions. Finally, conclusions and suggestions for future developments are provided in section \ref{sec5}.

\section{Data and Problem Setup}
\label{sec2}
%Text here ===>>>

\subsection{Spatiotemporal Variability of Sea Surface Temperature in the North Atlantic}
\label{sec21}
%Text here ===>>>
We consider the spatiotemporal variability of sea surface temperature
(SST) in the North Atlantic over the last 800 years of the
pre-industrial control simulation, or piControl, a simulation in which external
forcing is held fixed, from the Community Earth System Model (CESM)
\cite{danabasoglu2020community} as part of the sixth phase
of the Coupled Model Intercomparison Project (CMIP6). CESM2 is a
global coupled ocean-atmosphere-land-land ice model, and the piControl
simulation considered herein uses the Community Atmosphere Model (CAM6) and Parallel Ocean Program (POP2) at a nominal 1$^o$ horizontal
resolution in both the atmosphere and ocean. Readers can refer to \cite{danabasoglu2020community} for details.  These data are
publicly available from the CMIP archive at
https://esgf-node.llnl.gov/projects/cmip6 and its mirrors. These
monthly data display variability on a large range of spatial and
temporal scales. The largest spatial variation is in the meridional
(i.e., latitudinal) direction, while the largest temporal variation is at
the annual timescale and represents the seasonal cycle. Because both variations are easily learned and predicted, we preprocess the
data to remove these components. The latitudinal variation is
eliminated by subtracting the time-mean SST at each geographical
location, and the seasonal cycle is removed by considering a 12-month moving average also at each geographical location.

\subsection{Formulation of the Learning Problem}
\label{sec22}
%Text here ===>>>

\begin{figure}[H]
    \centering
    \includegraphics[width=0.9\textwidth]{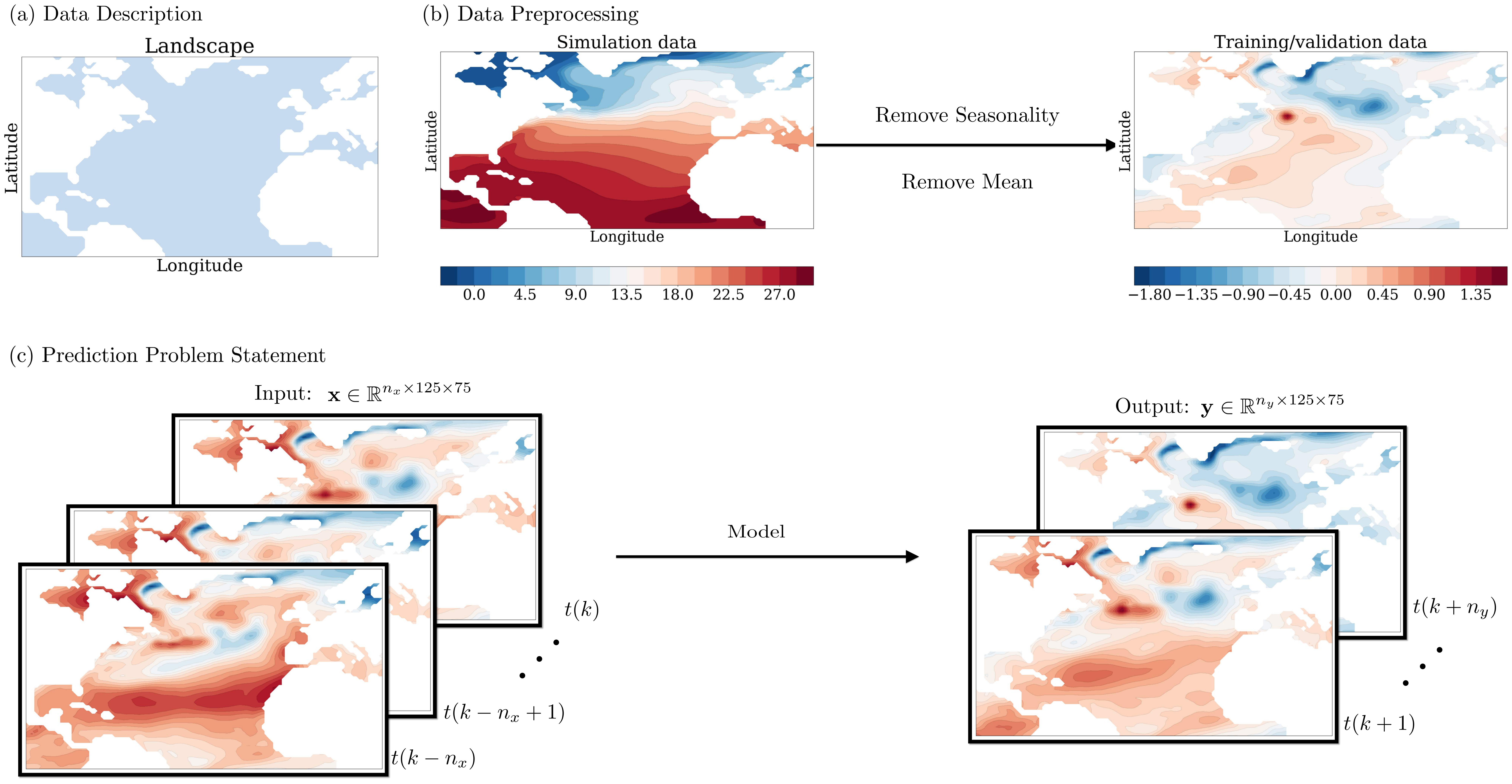}
    \caption{(a) The North Atlantic domain is shown in light blue. (b) Data preprocessing: The temporal mean and seasonal cycle are removed because they are easy to predict. While the original data span a range of $[-1.79, 30.41]$C, the interannual SST anomaly fields span a range of $[-1.68, 1.35]$C. The anomaly fields are obtained by removing the temporal mean and a mean seasonal cycle at each location. (c) Statement of the interannual SST anomaly prediction problem.}
    \label{fig:problem}
\end{figure}

Without loss of generality, we cast the near-term climate prediction problem in a video prediction format with the model input-output relationship described by a mapping of the form:
\begin{equation}
\label{eq1}
\mathcal{X} \in \mathbb{R}^{n_{x} \times H_{x} \times W_{x}}  \xrightarrow[]{f(\cdot)} \mathcal{Y} \in \mathbb{R}^{n_{y} \times H_{y} \times W_{y}},
\end{equation}
where $\mathcal{X}$ and $\mathcal{Y}$ denote the respective model input and output,  $n_{x}$ and $n_{y}$ represent samples  of $\mathbf{x}$ and $\mathbf{y}$ along the temporal dimension (that are chronologically ordered and at a constant sampling frequency), and SST is considered on a regular latitude (H) and longitude (W) spatial grid.
Equivalently, the prediction problem may be written as:
\begin{equation}
\label{eq2}
\mathbf{x}_{\mathbf{k}+\mathbf{n_y}}, \ldots, \mathbf{x}_{\mathbf{k}+\mathbf{2}}, \mathbf{x}_{\mathbf{k}+\mathbf{1}}=f \left(\mathbf{x}_{\mathbf{k}}, \mathbf{x}_{\mathbf{k-1}}, \ldots, \mathbf{x}_{\mathbf{k-n_x}}\right),
\end{equation}
where $\mathbf{x}_{\mathbf{k}}$ denotes the current state. Direct prediction of futures states $\mathcal{Y} = [\mathbf{x}_{\mathbf{k}+1}, \dots, \mathbf{x}_{\mathbf{k}+\mathbf{n_{y}}}]$ are made given a sequence of past and current states $\mathcal{X} = [\mathbf{x}_{\mathbf{k}-\mathbf{n_{x}}}, \dots, \mathbf{x}_{\mathbf{k}}]$. %Specifically, the problem centers on pixel-wise regression, which can be challenging considering the high dimensionality of the inputs and outputs $\mathcal{Y}$. %(e.g., in our case study $\mathcal{X} \in \mathbb{R}^{315000}$ and $\mathcal{Y} \in \mathbb{R}^{8750}$).
The schematic in Figure \ref{fig:problem} outlines the prediction problem.

\section{Methodology}
\label{sec3}
%Text here ===>>>

The goal is to develop efficient probabilistic deep learning models for near-term climate prediction while using advanced inference methods in the context of deep neural networks. This section describes our Bayesian learning strategy, the network architectures used, and other specifics regarding the training and testing procedure.

\subsection{Bayesian Deep Learning}
\label{sec31}
%Text here ===>>>

Estimation and quantification of the various sources of uncertainty is critical to establish the reliability of an ML model and provide an assessment of confidence in its predictions \cite{ghahramani2015probabilistic}.
This aspect of modeling is particularly important in the context of deep learning because of the large number of parameters that have to be leared in the DL setting.
To that end, we consider a probabilistic formulation that allows for characterizing uncertainties associated both with the data and model \cite{kendall2017uncertainties}. 

We assume that the weights have a probability density function of a fully factorized Gaussian prior with zero mean and a  precision  $\alpha$ that is Gamma-distributed.
Specifically, Bayesian deep learning (BDL) treats the network parameters $\mathbf{w}$ as random variables that can be generated via a prior distribution $p(\mathbf{w})$. By constructing the likelihood function $p(\mathcal{D}|\mathbf{w})$ from the given training data set $\mathcal{D}=\{\mathbf{x}_i, \mathbf{y}_i\}_{i=1}^{Ntrain}$, Bayes’ rule can be used to infer the posterior distribution of the network parameters $\mathbf{w}$:
\begin{equation}
\label{eq3}
p(\mathbf{w}|\mathcal{D}) = \frac{p(\mathcal{D}|\mathbf{w}) \, p(\mathbf{w})}{p(\mathcal{D})}.
\end{equation}
Subsequently, the predictive distribution $p(\mathbf{y}|\mathbf{x}) \equiv p(\mathbf{y}|\mathbf{w}; \mathbf{x})$ can be obtained by sampling the posterior $\mathbf{w} \sim p(\mathbf{w}|\mathcal{D})$.

%To better apply the BDL outlined above to the climate data, the following important properties should be observed: (1) \textbf{Hierarchical prior}: 

For the regression problem stated in Section \ref{sec22}, consider a deterministic neural network $\mathbf{y} = f (\mathbf{x}, \mathbf{w})$, where $\mathbf{x}$ is the input, $\mathbf{y}$ is the output and the parameters $\mathbf{w}$ include both the weights and biases.
While deterministic DL models treat the network parameters $\mathbf{w}$ as deterministic unknowns, BDL considers $\mathbf{w}$ as random variables to account for epistemic uncertainty induced both by limitations of the model (hypothesis set) considered and limited sampling of the data.
A further additive noise term $\mathbf{n}$ is used to model the irreducible aleatoric uncertainty in the data in this setting leading to 

\begin{equation}
\label{BDL}
\mathbf{y} = f (\mathbf{x}, \mathbf{w}) + \mathbf{n}.
\end{equation}

% {\bf Xihaier: What are you doing with $\mathbf{n}$ during testing???
%   Are you drawing from the posterior distribution of beta, computing
%   $\mathbf{n}$ and adding it to $f (\mathbf{x}, \mathbf{w})$ as in the above equation????}

\subsubsection{Prior definition}
\label{sec311}
As little is known about the network parameters before training, a non-informative prior typically is suggested to reduce and minimize the bias associated with the introduction of a prior \cite{neal2012bayesian}. Assuming that the prior is a fixed distribution independent of the input, we find imposing a sparsity-inducing prior on weights $\mathbf{w}$ via a hierarchical Bayesian model provides good performance. In particular, epistemic uncertainty of model parameters is described by a fully factorized Gaussian with zero mean and Gamma-distributed precision:
\begin{equation}
\label{eq: prior_1}
p(\mathbf{w}) = p(\mathbf{w} \mid \alpha)=\mathcal{N}\left(\mathbf{w} \mid 0, \alpha^{-1} \mathbf{I} \right), \quad p(\alpha)=\operatorname{Gamma}\left(\alpha \mid a_{0}, b_{0}\right)
\end{equation}
This results in a prior with a Student's T-distribution centered at zero. By tuning the rate parameter $a_{0}$ and the shape parameter $b_{0}$, one can employ a wider region with heavy tails than a standard Gaussian \cite{luo2020bayesian, zhu2018bayesian}. In this study, $a_{0} = 1$ and $b_{0} = 0.05$ are the values taken for the rate and shape parameters. On the other hand, aleatoric uncertainties capturing the noise in the data are assumed to be homoscedastic. Here, we prescribe additive noise $\mathbf{n}$ same for all output pixels/grids for lower memory and faster computation. Explicitly, the noise term is defined as $\mathbf{n} = \sigma \boldsymbol{\epsilon}$, where $\sigma$ is a scalar denoting the standard deviation of the data and $\boldsymbol{\epsilon}$ is Gaussian noise, i.e., $\boldsymbol{\epsilon} \sim \mathcal{N}(\boldsymbol{0}, \boldsymbol{I})$. In this work, the noise precision $\beta = 1/\sigma^2$ is modeled as a random variable with a conjugate prior in the form of
\begin{equation}
\label{eq: prior_2}
p(\beta) = \operatorname{Gamma}\left(\beta \mid a_{1}, b_{1}\right) 
\end{equation}
to better simulate the real-world applications. In most applications, the prior noise variance is assumed to be very small, e.g., $1 \times 10^{-6}$, so that providing a good initial guess for the prior hyperparameters \cite{gramacy2012cases}. Here, we consider the shape and rate parameters to be $a_{1} = 2$ and $b_{1} = 1 \times 10^{-6}$. It is worth noting that $\beta$ is a learnable parameter that is derived from data. Consequently, in the posterior estimation step, we will learn model parameters $\mathbf{w}$ and data parameter $\beta$. Unless otherwise specified, let $\boldsymbol{\theta} = \{ \mathbf{w}, \beta \}$ denote all the uncertain parameters for brevity. 

\subsubsection{Posterior estimation}
\label{sec312}
The second step of Bayesian learning is to estimate the posterior with predefined prior distributions. One of the standard ways to obtain the approximate posterior is to use sampling methods \cite{neal2012bayesian}. Given a large number of network parameters, e.g., tens or hundreds of millions in a modern deep learning model, this approach can be slow and difficult to converge. In recent years, significant progress has been made using VI methods as an alternative to approximate high-dimensional posterior distributions \cite{blei2017variational}.

Let $\mathcal{D} = \{ \mathbf{x}^{i}, \mathbf{y}^{i} \}_{i=1}^{N}$ be the training data. With a specified prior and a specified functional form for the likelihood, VI casts the Bayesian inference problem as an optimization problem. For a given likelihood $p\left(\mathbf{y} \mid \mathbf{x}, \boldsymbol{\theta}\right)$ and prior $p(\boldsymbol{\theta})$, the latter seeks to minimize the KL divergence between a proxy distribution $q (\boldsymbol{\theta})$ and the posterior distribution $p(\boldsymbol{\theta} \mid \mathcal{D})$:

\begin{equation}
\label{eq: KL}
\boldsymbol{\theta}^{*}=\underset{q \in \mathcal{Q}}{\arg \min } \mathbb{K} \mathbb{L} (q(\boldsymbol{\theta}) \| p(\boldsymbol{\theta} \mid \mathcal{D}))=\underset{q \in \mathcal{Q}}{\arg \min } \mathbb{E}_{q}[\log q(\boldsymbol{\theta})-\log \tilde{p}(\boldsymbol{\theta} \mid \mathcal{D})+\log Z]
\end{equation}

where $\tilde{p}(\boldsymbol{\theta} \mid \mathcal{D}) = \prod_{i=1}^{N} p\left(\mathbf{y}^{i} \mid \boldsymbol{\theta}, \mathbf{x}^{i}\right) p(\boldsymbol{\theta})$ is the unnormalized posterior and $Z = \int \tilde{p}(\boldsymbol{\theta}) d \boldsymbol{\theta}$ is the normalizer, also called model evidence. In practice, the normalization constant is not considered in the KL divergence minimization \cite{blei2017variational}. Also, the proxy distribution $q (\boldsymbol{\theta})$ is usually parameterized with a specified form of distributions $\mathcal{Q}$, inevitably introducing deterministic biases \cite{blundell2015weight}. In this work, SVGD, a nonparametric VI algorithm, is adopted \cite{liu2016stein}. Without defining a variational approximation family as parametric VI methods do, SVGD employs a set of independent identically distributed particles $\boldsymbol{\theta}_1, \boldsymbol{\theta}_2, \dots, \boldsymbol{\theta}_M$ and minimizes the KL divergence between the empirical measure of these particles and the true posterior. The central idea is to iteratively move the set of particles toward the true posterior using the gradient descent method:

\begin{equation}
\label{eq: iterative}
\boldsymbol{\theta}_{j}^{t+1}=\boldsymbol{\theta}_{j}^{t}+\epsilon_{t} \phi\left(\boldsymbol{\theta}_{j}^{t}\right),
\end{equation}
where $\epsilon$ is a small number representing the step size in the updating scheme and $\phi(\cdot)$ is the optimal perturbation direction that gives the steepest KL divergence gradient:

\begin{equation}
\label{eq: SVGD}
\phi(\boldsymbol{\theta})=\frac{1}{n} \sum_{j=1}^{n}[k\left(\boldsymbol{\theta}_{j}^{t}, \boldsymbol{\theta} \right) \underbrace{\nabla_{\boldsymbol{\theta}_{j}^{t}}\left(\log p\left(\boldsymbol{\theta}_{j}^{t}\right)+\log p\left(\mathcal{D} \mid \boldsymbol{\theta}_{j}^{t}\right)\right)}_{\text {gradient }}+\underbrace{\nabla_{\boldsymbol{\theta}_{j}^{t}} k\left(\boldsymbol{\theta}_{j}^{t}, \boldsymbol{\theta} \right)}_{\text {repulsive force }}]
\end{equation}
with $k(.,.)$ denoting a positive definite kernel. In this work, we choose a standard
radial basis function kernel for $k(.,.)$. In Equation (\ref{eq: SVGD}), the \textit{gradient} term pushes the particles toward high-density regions of the target distribution, and the \textit{repulsive force} term imposes diversity and prevents particle collapse \cite{liu2016stein}. Overall, the SVGD updating procedure can be summarized in five steps:

Step 1: compute the joint likelihood $\log p\left(\boldsymbol{\theta}_{t}^{j}\right)=\prod_{i=1}^{N} p\left(\mathbf{y}_{i} \mid \boldsymbol{\theta}_{t}^{j}, \mathbf{x}_{i}\right) p\left(\boldsymbol{\theta}_{t}^{j}\right)$. 

Step 2: calculate the gradient $\nabla_{\boldsymbol{\theta}_{t}^{j}} \log p\left(\boldsymbol{\theta}_{t}^{j}\right)$ by back propagation.

Step 3: compute the kernel matrix $\left[k\left(\boldsymbol{\theta}_{t}^{j}, \boldsymbol{\theta}_{t}^{i}\right)\right]_{i, j \in\{1, \cdots, M\}}$ and its gradient $\nabla_{\boldsymbol{\theta}_{t}^{j}} k\left(\boldsymbol{\theta}_{t}^{j}, \boldsymbol{\theta}_{t}^{i}\right)$.

Step 4: calculate the kernel Stein operator using equation (\ref{eq: SVGD}).

Step 5: Update $\boldsymbol{\theta}$ via stochastic gradient descent.

\subsubsection{Predictive distribution}
\label{sec313}

On completion of training, the model can be applied to make probabilistic predictions using unseen data samples $(\mathbf{x}^{*}, \mathbf{y}^{*})$; the predictive distribution is given by

\begin{equation}
\label{eq: predict}
p(\boldsymbol{y}^* \mid \boldsymbol{x}^*; \mathcal{D})=\int p(\boldsymbol{y}^* \mid \boldsymbol{x}^*; \boldsymbol{\theta}) \, \, p(\boldsymbol{\theta} \mid \mathcal{D}) d \boldsymbol{\theta}.
\end{equation}

Note that the SVGD algorithm provides a sample representation of the posterior $p(\boldsymbol{\theta} \mid \mathcal{D})$. Meaning one can use the learned SVGD particles to estimate the predictive distribution than the direct integration. Furthermore, effectively, the BDL model can be viewed as an ensemble of deep learning models where the number of ensemble members is given by the particle number $M$ (e.g., $M = 20$ in this work), we can approximate the moments of the predictive distribution using the Monte Carlo method. For instance, the ensemble mean prediction is given by the mean over the particles:

\begin{equation}
\label{eq: mean}
\mathbb{E}\left[\boldsymbol{y}^{*} \mid \boldsymbol{x}^{*}, \mathcal{D}\right]=\mathbb{E}_{p(\boldsymbol{\theta} \mid \mathcal{D})}\left[\mathbb{E}\left[\mathbf{y}^{*} \mid \mathbf{x}^{*}, \boldsymbol{\theta}\right]\right] \approx \frac{1}{M} \sum_{j=1}^{M} \mathbf{f}\left(\boldsymbol{x}^{*}, \boldsymbol{\theta}^{j}\right),
\end{equation}

and uncertainties are estimated utilizing the second moment of the predictive distribution:

\begin{equation}
\label{eq: cov}
\begin{split}
\operatorname{Cov} (\boldsymbol{y}^{*} \mid \boldsymbol{x}^{*}, \mathcal{D}) & = \mathbb{E}_{\boldsymbol{\theta}}\left[\operatorname{Cov}\left(\boldsymbol{y}^{*} \mid \boldsymbol{\theta},  \boldsymbol{x}^{*}\right)\right]+\operatorname{Cov}_{\boldsymbol{\theta}}\left(\mathbb{E}\left[\boldsymbol{y}^{*} \mid \boldsymbol{\theta}, \boldsymbol{x}^{*}\right]\right) \\
& \approx \frac{1}{M} \sum_{j=1}^{M}\left(\left(\mathbf{n}_j\right)^{-1} \mathbf{I}+\mathbf{f}\left(\boldsymbol{x}^{*}, \boldsymbol{\theta}_j\right) \mathbf{f}^{\top}\left(\boldsymbol{x}^{*}, \boldsymbol{\theta}_j\right)\right) -\left(\frac{1}{M} \sum_{j=1}^{M} \mathbf{f}\left(\boldsymbol{x}^{*}, \boldsymbol{\theta}_j\right)\right)\left(\frac{1}{M} \sum_{j=1}^{M} \mathbf{f}\left(\boldsymbol{x}^{*}, \boldsymbol{\theta}_j\right)\right)^{\top}
\end{split}.
\end{equation}

With the estimated mean and variance, we can make the prediction for new data samples and quantify its associated predictive uncertainty.

\subsection{Architecture Design}
\label{sec32}
Even as the success of applying deep learning to problems in science and engineering depends crucially on the choice of network architecture, designing efficient and effective networks remains problem-specific and requires human expertise.
In this context, we note climate-relevant data typically are \textit{high dimensional}, \textit{geographically heterogeneous}, and most often result from {\em dynamical and other physical interactions over a diverse range of spatial and temporal scales} \cite{reichstein2019deep}. 

In regard to the high-dimensional nature of climate data, recent studies reveal that the intrinsic dimension captured by dimensionality reduction techniques tends to be too low and, therefore, insufficient \cite{kashinath2021physics}.
Consequently, rather than rely on dimensionality reduction techniques, we use an architecture that considers the full extent of the spatial degrees of freedom present in the data \cite{xu2021feature}.
Next, motivated by the fact that common yet important fluid-dynamic processes, such as advection and diffusion, are represented by regular stencils in the numerical solution of partial differential solutions governing the climate system, we use convolution layers as an integral aspect of the network.
Finally, to permit the learning of multiscale interactions, e.g., both local and remote interactions, we employ a bottleneck of sufficiently high dimension with additional optional fully connected layers in the bottleneck.
Notably, this design maintains the deep learning promise of automatically extracting features, allowing them to interact appropriately for the task on hand and subsequently projecting them back at the required resolution in an end-to-end fashion.

\begin{figure}[H]
    \centering
    \includegraphics[width=0.95\textwidth]{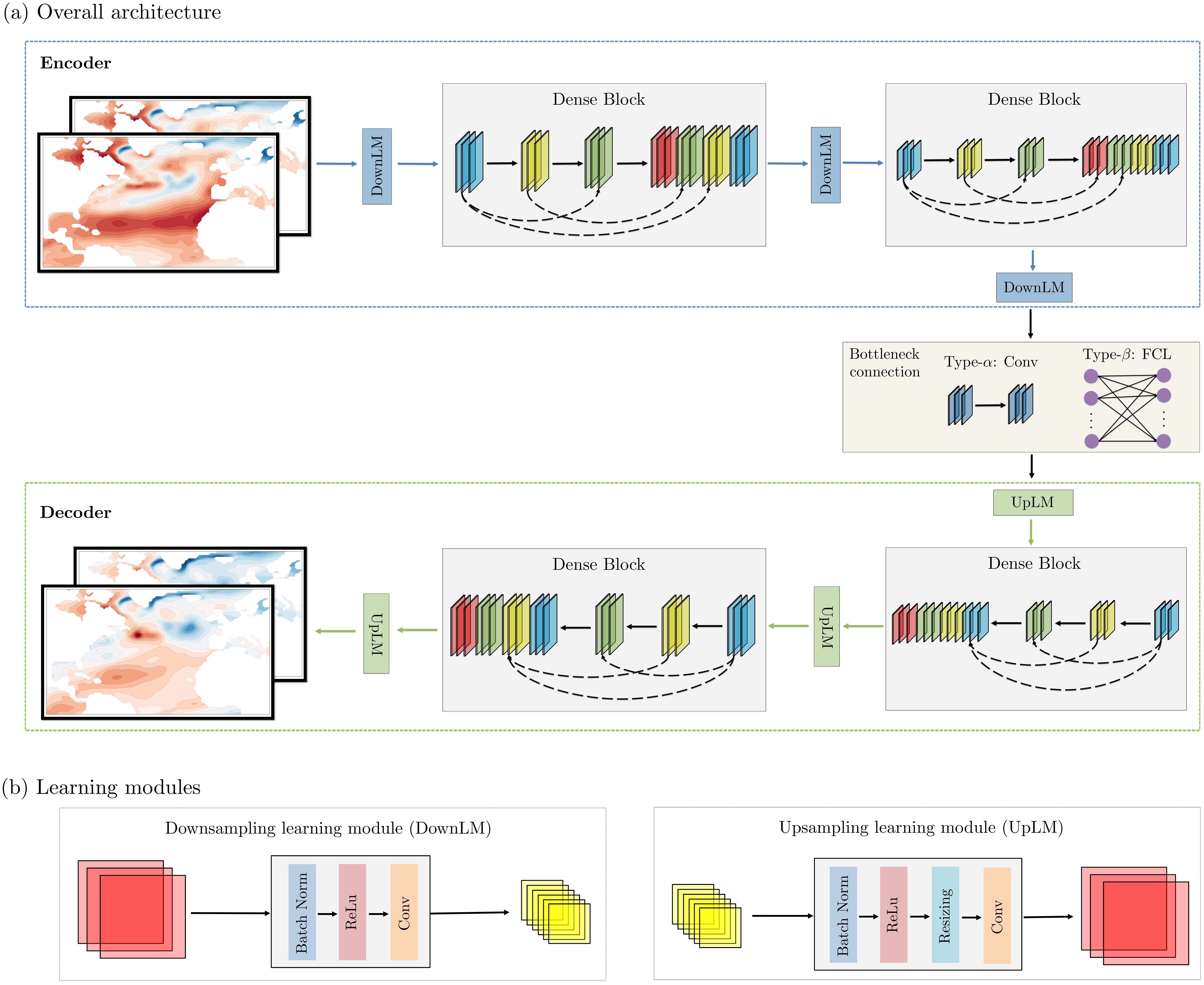}
    \caption{Densely connected convolutional neural networks-based encoder-decoder architecture.}
    \label{fig:model}
\end{figure}

In Figure \ref{fig:model}, the down- and up-sampling learning modules greatly reduce the number of network parameters, thereby accelerating the training process. Specifically, convolution operations are performed to reduce the data size and extract features. Non-adjacent connections then are established for aggregating extracted features \cite{he2016deep}. As most deep learning models are data-intensive, a densely connected convolutional network structure, known as \textit{dense block}, is adopted in our encoder-decoder architecture to reduce network parameters and support more stable learning \cite{huang2017densely}. Consequently, each layer can reuse the features extracted from all preceding layers in the dense block. Inside each dense block, a layer is defined as a set of composition operations usually denoted as \textit{convolution}, \textit{nonlinear} \textit{activation}, \textit{batch normalization}, and \textit{dropout} \cite{he2016identity}. We combine image resizing techniques and convolution in the upsampling learning module. In particular, transposed convolutions are commonly performed for upsampling the extracted features to the desired spatial dimensions. However, \citeA{odena2016deconvolution} find transposed convolutions with uneven overlapping cause a checkerboard pattern of artifacts. Therefore, image resizing techniques, such as nearest-neighbor or bilinear interpolation, serve as good alternatives. For instance, bilinear interpolation discourages high-frequency artifacts via an implicitly weighting filter, which is adopted here. Lastly, we observe pooling operations, usually implemented in-between successive convolution layers to reduce the size of feature maps, can deteriorate prediction performance. Knowing climate data, by nature, differ from most computer science application data (e.g., handwriting digits), we argue that max or average pooling may lead to the loss of distinctive features to infer finer pixel-wise regression. Hence, pooling operations are not considered. Instead, a convolution operator with a non-unit stride is used to manage the feature sizes \cite{dumoulin2016guide}.

For the multiscale interactions that typically underlie climate data, a brute force solution consists of a fully connected multilayer perceptron (MLP) model \cite{goodfellow2016deep}. Thus, the influence from long-distance locations is seamlessly integrated into the MLP architecture. However, an MLP model can be memory-demanding and computationally prohibitive because the number of total parameters increases too quickly, i.e., as the cumulative product of the number of perceptrons in each layer. A better, more efficient way to account for long-distance effects is to add a fully connected linear layer in the feature space, ideally at the bottleneck level. The combination of convolutional and fully connected layers has shown its effectiveness in many computer vision tasks, such as objective detection and image segmentation \cite{dosovitskiy2020image, rasp2020weatherbench}. Hence, we also consider fully connected layers between the densely connected encoder and decoder. In such a case, it is anticipated that the fully connected layers will relieve part of the burden on the encoder-decoder parts of the network to learn remote interactions, freeing them to better represent spatiotemporally local interactions---and convolution layers excel at these tasks.

Following such reasoning, we develop our deep learning \texttt{(DL)} model and its Bayesian version (\texttt{BDL}) for climate prediction. We also include the state-of-the-art dynamics forecasting model, \texttt{ConvLSTM}, in the study for better comparison. Here, all three models have a convolutional encoder to condense information. Of note, \texttt{DL} and \texttt{BDL} share the same architecture, and \texttt{BDL} is initialized and stored in a predefined particle number. We defer the network parameter details, including size of the convolving kernel, stride of the convolution, zero-padding size, etc., to \ref{app1}. Figure \ref{fig:model} offers a graphic illustration of the proposed model.

% two baseline models for climate prediction (\texttt{DL}-$\alpha$ and \texttt{DL}-$\beta$). We further build their Bayesian version for comparison (\texttt{BDL}-$\alpha$ and \texttt{BDL}-$\beta$). Also, we include the state-of-the-art dynamics forecasting model, \texttt{ConvLSTM}, in the study.

% \begin{itemize}
%     \item \texttt{DL}-$\alpha$: a 20-layer dense encoder-decoder network only with convolutional layers adapted dynamics forecasting.
%     \item \texttt{DL}-$\beta$: a 20-layer dense encoder-decoder network with a fully connected linear layer at the bottleneck level.
%     \item \texttt{BDL}-$\alpha$: Bayesian version of \texttt{DL}-$\alpha$.
%     \item \texttt{BDL}-$\beta$: Bayesian version of \texttt{DL}-$\beta$.
%     \item \texttt{ConvLSTM}: a 3-layer Convolutional LSTM model used for spatiotemporal precipitation nowcasting.
% \end{itemize}

\subsection{Network Training}
\label{sec33}
%Text here ===>>>

The goal of network training is to minimize the mismatch between a prediction $\hat{\mathbf{y}} = f(\mathbf{x})$ and the correct output $\mathbf{y}$. For \texttt{DL} and \texttt{ConvLSTM}, $\hat{\mathbf{y}}$ denotes the model output. On the other hand, $\hat{\mathbf{y}}$ is defined as the predictive mean of Bayesian particles in \texttt{BDL}. The mean squared error (MSE) is selected as the criterion here to measure the difference between $\hat{\mathbf{y}}$ and $\mathbf{y}$ in each grid. Training details for all models are given as: (1) the data set is split into a training set consisting of $1280$ paired samples ($\{\mathbf{x}_i, \mathbf{y}_i\}_{i=1}^{1280}$) and a test set containing $128$ samples in all experiments; (2) the data is standardized by removing the mean and scaling to unit variance; (3) the batch size of both training and test sets are set the same as $128$ for non-probabilistic models and $32$ for Bayesian models; (4) the Adam stochastic gradient descent algorithm is used as the default optimizer with weight decay specified to $5 \times 10^{-4}$ to regularize the weights via an L2 penalty \cite{kingma2014adam}, which ensures the model generalizes better to unseen data; (5) the initial learning rate is set to $0.001$ with a dynamic scheduler to reduce the learning rate by a factor of 10 when the computed metric has stopped improving; and (6) the dropout technique is used after each convolutional layer to further reduce overfitting and improve generalization error \cite{hinton2012improving}, where the probability of an element to be zeroed is set to $0.5$.

\subsection{Implementation Notes}
\label{sec34}
%Text here ===>>>

We consider Python 3.7.3 and PyTorch 1.7.0 to implement the methodology. All data processing and model training assessments are carried out on a NVIDIA Tesla V100 graphics processing unit (GPU) card with 16 GB high-bandwidth memory (HBM). For full reproducibility of the results, the computer codes and data can be found at \url{https://xihaier.github.io/} upon publication.

\section{Results and Discussions}
\label{sec4}

\begin{figure}[H]
    \centering
    \includegraphics[width=0.7\textwidth]{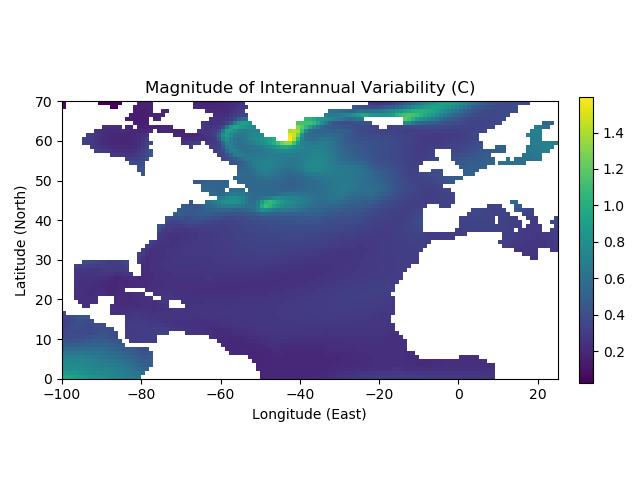}
    \caption{The standard deviation of the interannual anomaly of
SST is plotted here as a measure of the magnitude of interannual
variability (in units of degree Centigrade).}
    \label{fig:std}
\end{figure}

Figure~\ref{fig:std} shows a measure of the magnitude of interannual
internal/natural variability that we are interested in predicting. It
is computed as the standard deviation of the interannual anomaly of
SST. We also refer to this as the climatological standard deviation.
This variability is seen to be
geographically heterogeneous with the largest variations in the
subpolar North Atlantic and the isolated part of the Eastern Pacific
(related to ENSO).
We nondimensionalize prediction error using the climatological standard
deviation to make the errors geographically commensurate and to facilitate
comparison of prediction error across different regions.

\begin{figure}[H]
    \centering
    \includegraphics[width=0.95\textwidth]{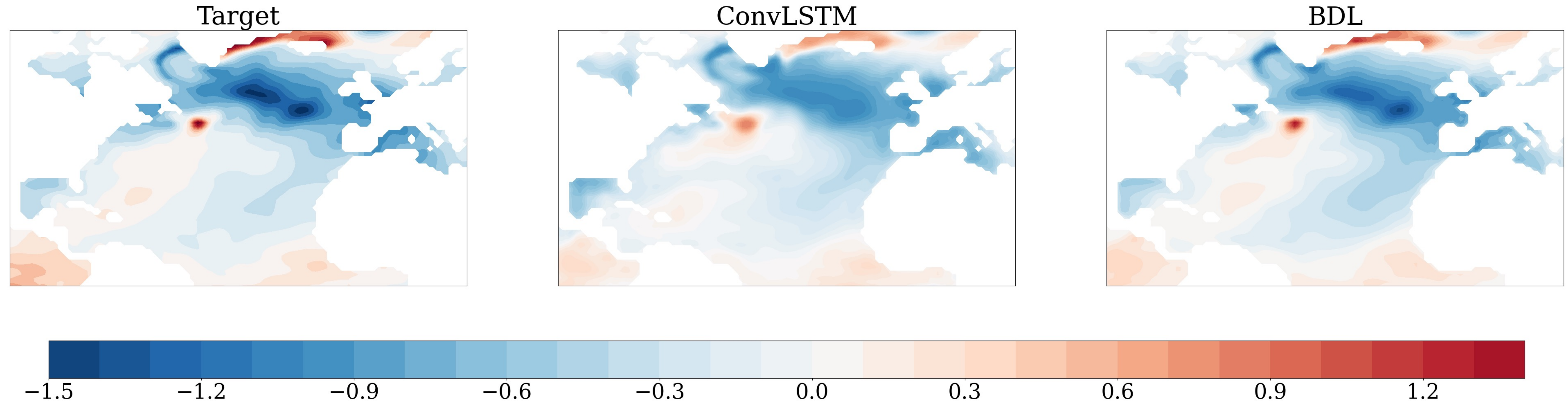}
    \caption{Prediction results of \texttt{BDL} and \texttt{ConvLSTM} for a randomly selected test sample at a lead time of six months. While both predictions correlate reasonably with the target, the BDL predictions are seen to sharper, even while exhibiting lower errors.}
    \label{fig:instance}
\end{figure}

Figure~\ref{fig:instance} compares the predictions of SST in the North
Atlantic at a lead time of six months for a randomly selected test
case. The target spatial distribution of SST is shown in the left
panel and we recall that both the strong, time-mean latitudinal
variability and the seasonal cycle have been removed in the current
study so as to focus largely on the interannual component of
variability. As such the main variability seen in the target
distribution is related to the spatial heterogenity of the nature of
interannual variability. On comparing the predictions in the other two
panels with the target distribution, it is seen that both the convLSTM
prediction (center panel) and the ensemble-mean of the BDL prediction
(right panel) successfully capture the main aspects of the spatial
distribution such as the warm spot near Grand Banks, the cooler
temperatures of the subpolar gyre, the warm anomaly in the East
Greenland current, the lower variability in the subtropical gyre
region, and others. However, it is also seen that whereas the convLSTM
predictions tend to be more diffuse, features in the BDL prediction
are better correlated with the target and tend to be sharper even
though we are considering an ensemble-average. This is suggestive of
better performance of the probablisitic BDL system as compared to the
deterministic convLSTM system \cite{xingjian2015convolutional,
  parkmachine, xu2021feature}.

\begin{figure}[H]
    \centering
    \includegraphics[width=0.9\textwidth]{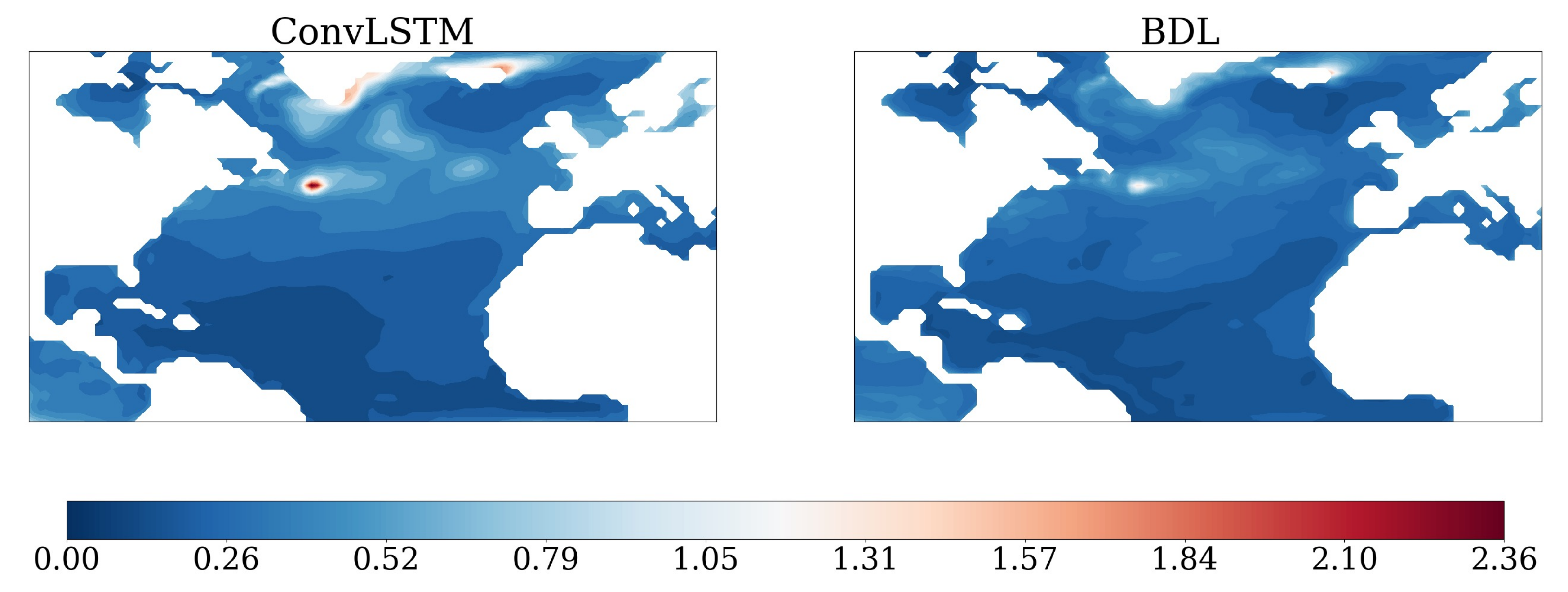}
    \caption{Prediction error map showing the non-dimensional root mean square error at a lead time of six months. The BDL errors are seen to be lower.}
    \label{fig:NDRMSE_2d}
\end{figure}

This observation is confirmed on examining the prediction accuracy averaged over the entire test data set of 128 test samples. Figure \ref{fig:NDRMSE_2d} shows the prediction error pattern map at a lead time of six months for convLSTM and BDL. The error is specifically defined as the non-dimensional root mean square error (NDRMSE), with the climatological standard deviation being used to non-dimensionalize the RMSE at each location. Results at other lead times are qualitatively similar. 

\begin{figure}[H]
    \centering
    \includegraphics[width=0.9\textwidth]{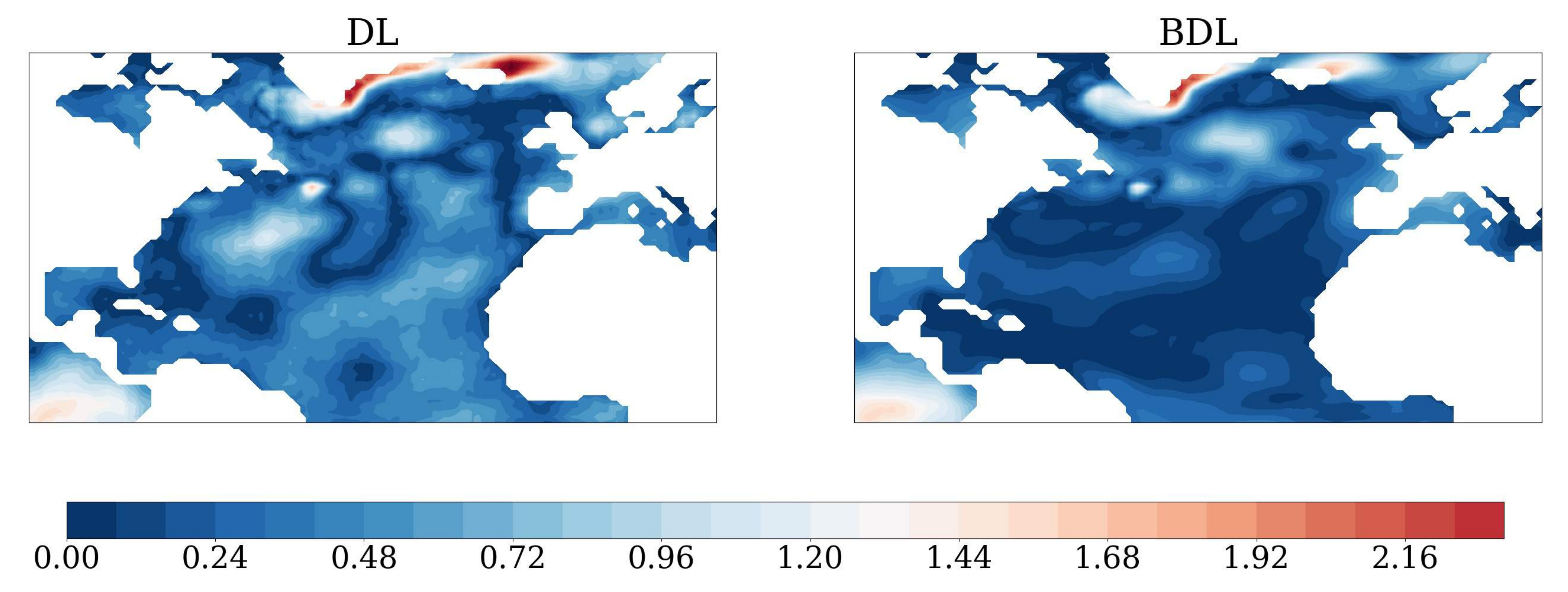}
    \caption{A comparison of errors between the deterministic and
      Bayesian predictions at a prediction lead time of a year. The
      ensemble mean BDL prediction is seen to have smaller errors than
      its deterministic counterpart.}
    \label{fig:overfitting}
\end{figure}

Figure \ref{fig:overfitting} compares the prediction error of \texttt{DL} and \texttt{BDL} for the same test sample as shown in Fig.~\ref{fig:instance}, but at a lead time of twelve months. Two features are seen in this comparison:  error is seen to be lower in the Bayesian model, and the error in the deterministic model is seen to have smaller scale features. This suggests the possibility that the deterministic DL model is more prone to overfitting and that the averaging inherent in the Bayesian DL acts to regularize the BDL predictions. This, in turn, reiterates  the need to assess model and data uncertainty using probabilistic modeling techniques, particularly when considering deep networks \cite{ghahramani2015probabilistic}.

\begin{figure}[H]
    \centering
    \includegraphics[width=0.75\textwidth]{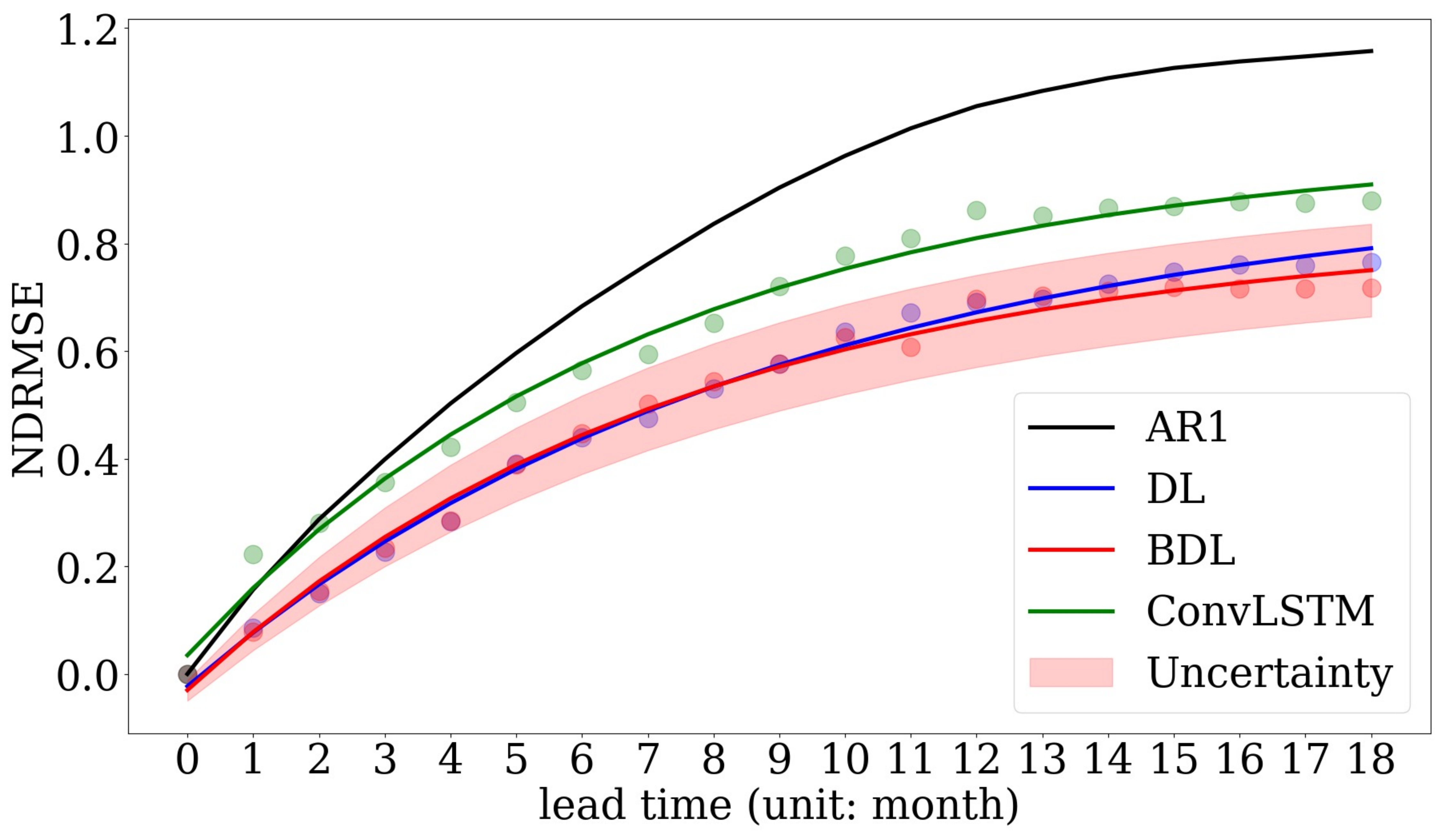}
    \caption{Domain averaged, and test set-averaged (root mean square)
      non-dimensional error as a function of prediction lead time for
      the various methods considered. The baseline model ``AR1''
      (first order auto regressive process) corresponds to damped
      persistence. All ML models are seen to be better than the
      baseline. The deterministic and Bayesian versions of the
      convolution based FNN are seen to be better than convLSTM, a
      sophisticated RNN architecture. The Bayesian version of the FNN
      is slightly better than its deterministic counterpart, while
      simultaneously providing a measure of uncertainty in the
      prediction.}
% {\bf Xihaier: For consistency with Figs. 5 and 6,
%         each one of the error curves and the envelopes have to be
%         computed as the square root of the average over domain and
%         test-set of the nondimensional error. The AR1 error values are
%       indicated here as a comment}
    % array([0.15661651, 0.28712943, 0.3985741 , 0.5025135 , 0.59627536,
    %     0.6833441 , 0.76125082, 0.83605356, 0.9035376 , 0.9627359 ,
    %     1.01387267, 1.05503717, 1.08336539, 1.10721883, 1.12605396,
    %     1.13806286, 1.14744734, 1.15741604])
    \label{fig:NDRMSE}
\end{figure}

Figure \ref{fig:NDRMSE} compares the domain averaged error as a
function of prediction lead time in the various models considered.
For convLSTM, DL, and BDL, NDRMSE averaged (over the domain and) over
the test sets are shown in filled circles.  A further fit of the
inidividual points using an function of the form
$\text{NDRMSE} = \alpha \left( 1 -\exp (-\beta* \tau)\right)$, with
$\tau$ denoting the prediction lead time is also shown.  The fit is
obtained by minimizing the residual in a nonlinear least-squares
problem using a trust-region algorithm \cite{conn2000trust}.  Along
with the previously mentioned models, the damped persistence fit,
obtained by fitting a first order autoregressive model is shown and
indicated as \texttt{AR1}.

First, it is seen that for the most part, irrespective of whether it
is FNNs or RNNs, and whether it is deterministic ML models or
probabilistic ones, each of the ML models performs better than damped
persistence.  Next, it is interesting to note that \texttt{DL} and
\texttt{BDL}, both feedforward networks (FNN) outperform the convLSTM
model, a recurrent network (RNN) that has previously been seen to
provide good performance in a variety of temporal prediction settings.
Furthermore, the probabilistic Bayesian deep learning model performs
better than its deterministic counterpart, as discussed earlier.
Finally, for BDL, uncertainty is estimated as the standard deviation
of the ensemble spread and is shown by the red envelope in Figure
\ref{fig:NDRMSE}.  The uncertainty in prediction is seen to increase
with lead time.  In an RNN setting, in contrast to the current FNN
setting, it is typical to train the recurrent network to produce a one
step prediction. Thereafter, predictions at longer lead times are
produced based on both the input at the current time and output at the
previous time. Thus, compounding of error and uncertainty with
increasing lead time explains the increase of both error and
uncertainty with increasing lead time in an RNN setting. On the other
hand, in the current FNN setting, the straight forward process of
compounding of error and uncertainty with increasing lead time is
absent and thus constitutes an independent validation of the current
FNN approach. Indeed, we go on to consider the nature of these
increases with time further in the following section.

% {\bf We need to discuss this: As previously noted, this could be due to the nonlinear relationship between forecast lead time and probable outcome scenarios. In the presence of such data uncertainty, a deterministic model trained to minimize the mean squared error will tend to predict the mean of the distribution of probable outcomes, but a probabilistic model (\texttt{BDL}) can account for data variability.}

In terms of computational cost, our implementation of the \texttt{DL}
model (primarily based on convolution operations) is approximately 20
times faster than the \texttt{ConvLSTM} model.  We have further
verified this speed-up in the contest of a larger data set
\cite{parkmachine, xu2021feature}.  It is worth noting that the time
complexity of training a Bayesian network is theoretically $O(n)$,
where n denotes the number of particles in the SVGD algorithm
\cite{liu2016stein}.  In all experiments, we find that the
\texttt{BDL} model achieves better scalability than linear complexity
and requires less training time than the \texttt{ConvLSTM} model.

\subsection{Quantification and Nature of Uncertainty as Represented in the Bayesian Deep Learning Model }
\label{sec42}
%Text here ===>>>

Now that we have a probabilistic prediction system that takes into
account the possibility of a range of models fitting the training data
in a Bayesian framework, we are able to generate a range of outcomes
for the test data as well. Such a distribution of predictions has
multiple uses including obtaining information of alternative future
evolutions and the possibility of predicting extreme events. However,
given the experimental nature of the probablistic prediction system
considered, we presently confine ourselves to examining the quality of
the predictions and its utility in providing insights into the
methodology itself.

In addition to the slight improvement in prediction skill when
compared to deterministic deep learning models (\texttt{DL} and
\texttt{ConvLSTM}), \texttt{BDL} allows for a means of quantifying the
uncertainty inherent in the data and model. For example, in the
particular approach we consider, learning the posterior distribution
of $\beta$, a parameter related to data uncertainty and whose prior
distribution is given in (\ref{eq: prior_2}) serves to quantify data
uncertainty. Likewise, the learning of the posterior distribution of
the model parameters $\mathbf{w}$, the priors for which are given in
(\ref{eq: prior_1}) serves to capture uncertainty in the model
itself. While we have already shown and briefly discussed the behavior
of uncertainty in the BDL in the previous section, we seek to analyze
it further in this section and see what further insight it may yield
into the workings of the Bayesian model.

% {\bf Xihaier, please send me a plot of the posterior distribution of
%   beta. In fact, if you can get to the likelihood, it would be nice
%   to plot all three: prior, likelihood, and posterior}

\begin{figure}[H]
    \centering
    \includegraphics[width=0.9\textwidth]{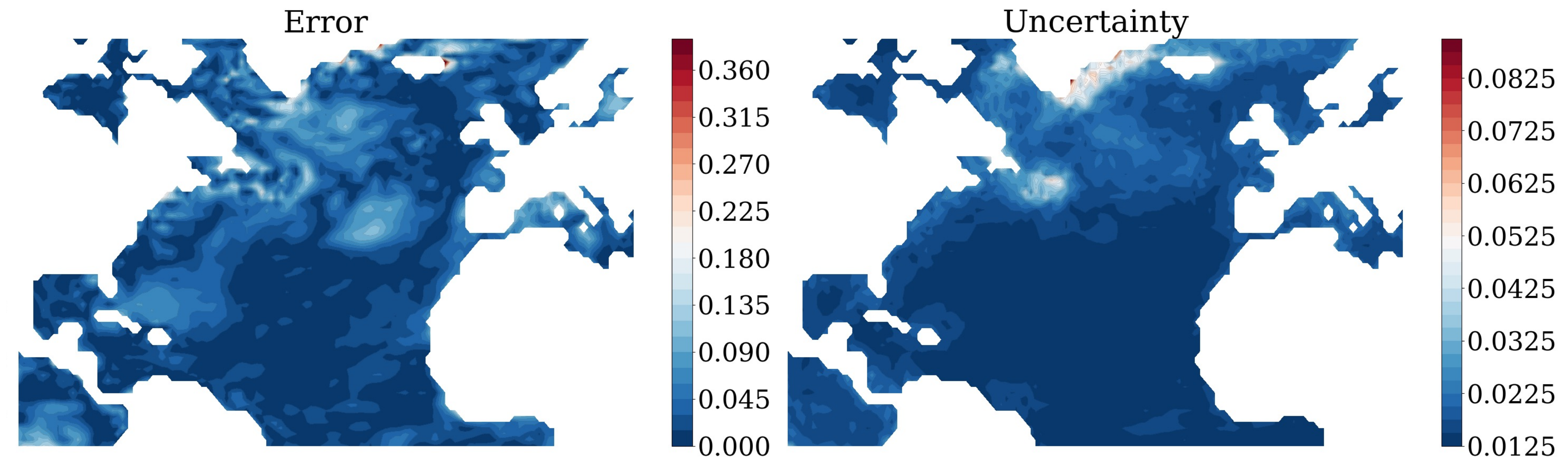}
    \caption{Error and uncertainty captured by the BDL model for the randomly selected sample presented in Fig. \ref{fig:instance}}
    \label{fig:f6}
\end{figure}

The rightmost panel in Figure \ref{fig:f6}, shows an estimate of
uncertainty in the prediction of the SST for a particular instance,
defined as the standard deviation of the ensemble.  Prediction
uncertainty varies depending on the underlying dynamical state and the
dynamics underlying SST varies significantly from the tropics to the
midlatitude and sub-polar regions.  This is reflected in the spatial
heterogenity of the uncertainty estimated in the BDL scheme. The
heterogenity of the estimated uncertainty is in agreement with the
hetrogenity of the magnitude of interannual variability shown in
Fig.~\ref{fig:std}. Finally, the figure also shows that the higher
(lower) level of error in the subpolar (tropical) region is
accompanied by a higher (lower) level of uncertainty.

\begin{figure}[H]
    \centering
    \includegraphics[width=0.95\textwidth]{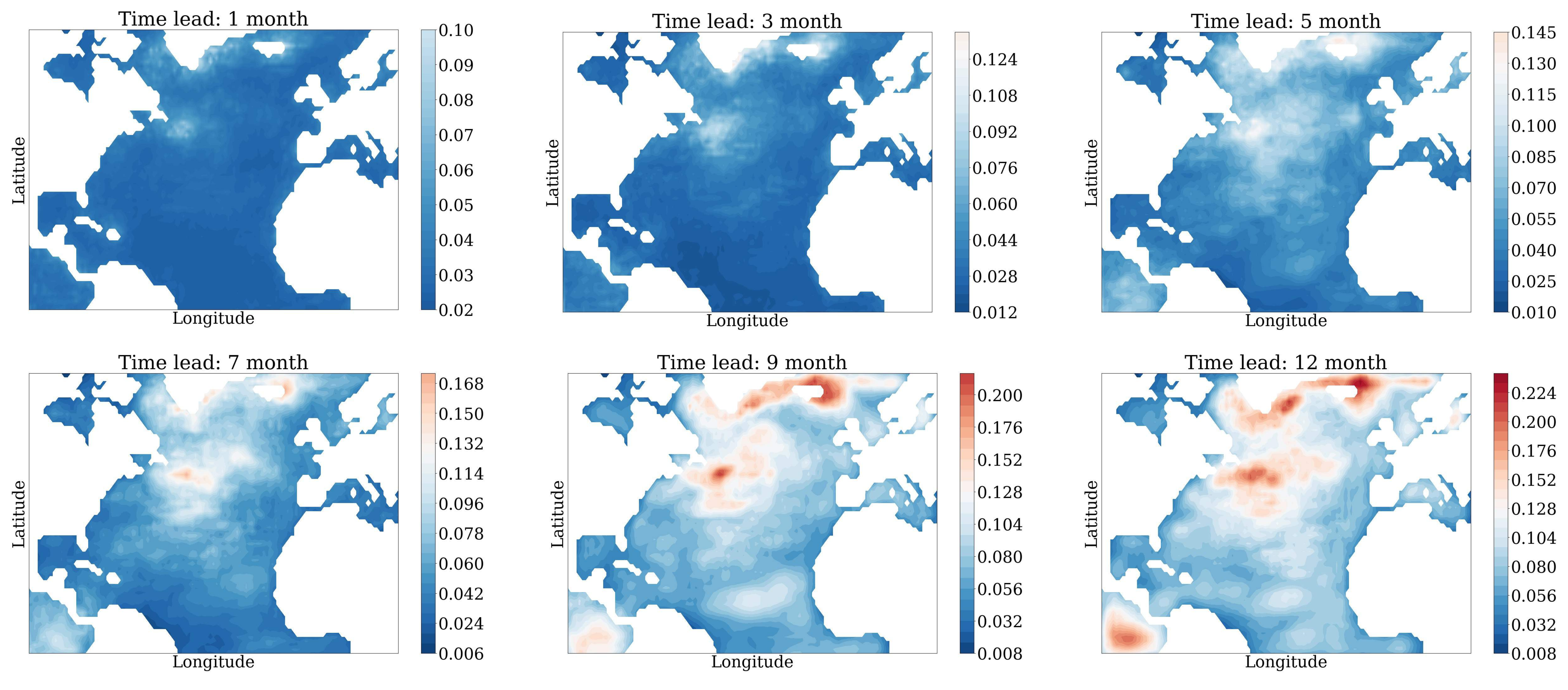}
    \caption{Prediction uncertainty.}
    \label{fig:snap_uncer}
\end{figure}

We previously saw the growth of uncertainty with increasing prediction
lead time in a domain-averaged sense in Fig.~\ref{fig:NDRMSE}. Figure
\ref{fig:snap_uncer} shows the spatial distribution of the growth of
uncertainty with prediction lead time. The spatial heterogenity is
related to the hetrogenity of the dynamics governing the evolution of
SST as discussed previously.  The higher level of uncertainty in the
isolated patch of the Pacific in the southwest corner of the domain is
likely due to the fact that the dynamics in that region is controlled
more by processes in the rest of the Pacific that is not considered
presently.

Next, we focus attention on the nature of the relationship between
prediction error and uncertainty. The reason for doing this is because
in a realistic situation we do not have verification data. As such, we
cannot directly estimate error in (or skill of) the predictions. So,
the question is, as {\em to what extent the uncertainty in the
  prediction can be used as a measure of prediction error?} It is
somewhat natural to think that the future state will be close to the
ensemble mean when the dispersion of the ensemble is small and
conversely for the accuracy of the ensemble mean to small when the
ensemble is highly dispersed. As such, we proceed to quantify the
relationship between ensemble spread and prediction accuracy of BDL
when verification data is present.

\begin{figure}[H]
    \centering
    \includegraphics[width=0.95\textwidth]{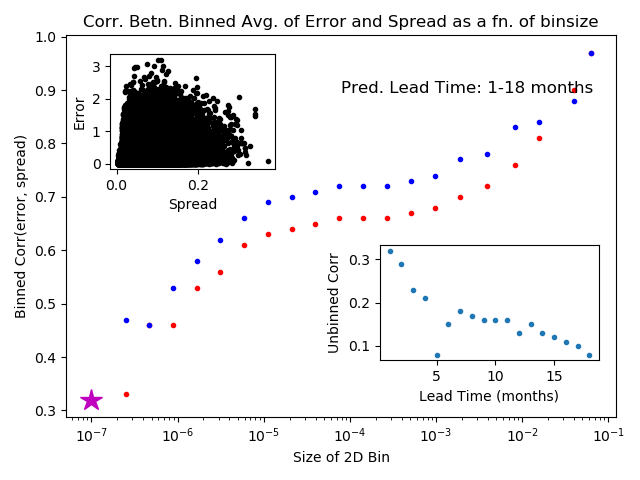}
    \caption{Analysis of the spread-error relationship for BDL. The
      inset at the top-left shows a scatter-plot of error vs. spread
      over the full period of prediction (1-18 month lead time). The
      data is highly dispersed and the Spearman correlation
      coefficient is a modest 0.33. The inset in the bottom-right
      shows a plot of the correlation coefficient as a function of
      prediction lead time. The correlation is seen to decay largely
      monotonically with increasing lead time. The main panel shows
      the correlation for a wide range of bin sizes and when outliers
      are eliminated; the blue and red dots correspond to two differnt
      thresholds for determining outliers. A distinct plateau of
      correlation is seen over a wide range of intermediate bin
      sizes. This analysis suggests that the uncertainty information
      from BDL sytem can be used to estimate prediction error to a
      certain degree.}
    \label{fig:f9}
\end{figure}

The inset in the top left panel of Fig. \ref{fig:f9} shows a scatterplot of error
of the ensemble-mean against ensemble-spread at each geographical
location, for each of the test instances, and at each prediction lead
time of between one and eighteen months. In this inset plot, a large degree of scatter is
seen and error and spread are moderately correlated; the Spearman
correlation coefficient is 0.33. (We prefer to use the rank-based
Spearman correlation coefficient since a) it is a nonparametric
measure of monotonicity of the relationship between the two variables,
b) unlike the Pearson correlation,  it does not assume that the
variables are normally distributed, which makes it more robust.) Here
we note that even in idealized 
experiments where the prediction model is perfect (in the sense that
it does not have any biases), for statistical reasons, the
spread-error correlation need not be large (e.g., see Barker91,
Houtekamer93).

The inset in the bottom-right shows the Spearman correlation
coefficient as a function of the prediction lead time. The correlation
coefficient is seen to decrease largely monotonically with lead
time. The low correlation at long lead time corresponds to the
prediction reverting to climatology.

The large scatter in the spread-error plot is due to considering the
predictions at the highest level of detail available. Given the usual
reduced predictability of the smaller of spatio-temporal scales
(equivalenty, increased predictability of the larger of spatiotemporal
scales), the question naturally arises as to the nature of this
correlation when the predictions are considered in a less-detailed or
more-aggregated fashion.  To examine this, we consider a bin-averaging
strategy: Previously, (e.g., \cite{wangbishop2003}) binning only along the
spread axis, followed by bin-averaging spread and error has been
suggested. However, on implementing this procedure, we find that it
leads to the predicted error falling in a range that is very narrow
compared to the actual range of errors. For this reason, we consider
binning along both the spread and error axes. First, the bin-edges for
each axis was determined such that each bin contains an equal number
of points. Next, the (spread, error) tuples were binned in the sense
of a two-dimensional histogram using the previously determined
bin-edges and spread and error were bin-averaged (RMS).  Next, the
least populated bins were eliminated and the correlation coefficient
was computed. In the main plot in Fig.~\ref{fig:f9}, the correlation
is plotted as a function of bin size, where for computing the size of
the bin, the range of values of both error and spread were set to
unity for simplicity. (This way, the inverse of the bin size gives the
number of bins.) For the points in blue, about a quarter of the points
were eliminated, while for the points in red, just less than a third
($\approx$ 31\%) were eliminated.  With this procedure, the range of
predicted error values is much closer to the actual range of error
values, and the plateau of correlation seen over a wide range of
intermediate bin sizes suggests that the the ensemble spread may be
used to estimate prediction error to a certain extent.
% Nevertheless there are
% other indicators that suggest that ensemble predictions with an FNN
% have characteristics that would be considered anomalous from the
% perspective of a physics-based prediction system. For example, the
% spread-error correlation is not seen to decay monotonically with lead
% time, and there is a rather large variation in the slope of
% spread-error relationship.

\begin{figure}[H]
    \centering
    \includegraphics[width=0.55\textwidth]{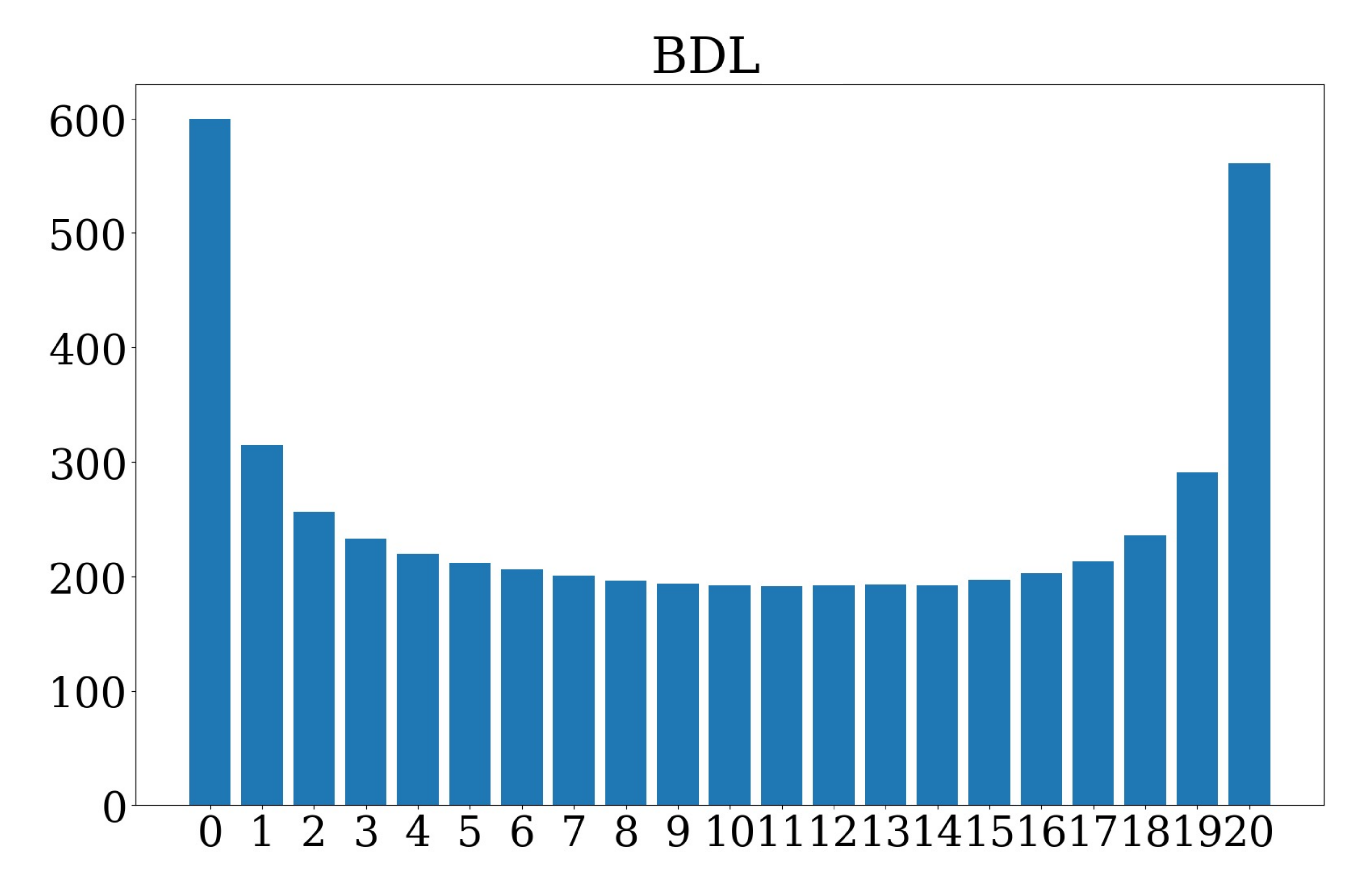}
    \caption{Rank histograms.}
    \label{fig:rankHist}
\end{figure}

Finally, we consider the verification rank histogram as a means for
characterizing the dependability and consistency of the BDL ensemble
\cite{hamill2001interpretation}. For each test sample, we have 20
predictions from the Bayesian surrogate. For each ocean point, for
each test instance and for each prediction lead time, we first rank
the predicted values, resulting in a vector of 20 scalars. We then use
the bisection algorithm to find the insertion position for the target
value in this vector. This is the rank of the target for that
particular ocean point in the particular test instance and at a
particular lead time. Figure \ref{fig:rankHist} shows the histogram of
the computed ranks.  For an optimal ensemble, the rank-histogram would
be flat. From the shape of the overall rank-histogram, it is seen that
the target value fall outside of the ensemble more often than in an
optimal ensemble, suggesting that the ensemble is slightly
under-dispersive. As such, we attempted to improve the nature of the
rank-histogram by increasing the number of particles, etc. However,
this effort was unsuccessful.  We therefore hypothesized that the
dispersivity of the ensemble was geographically heterogeneous given
the heterogenity of the interannual variability (Fig.~\ref{fig:std})
and the heterogenity of the estimated uncertainty (Fig.~\ref{fig:f9}). To
examine this, we present three heatmaps showing the counts of elements
contained in the left exterior $(\# 0$), interior ($\# 1 \sim 19$),
and right exterior ($\# 20$) bins, respectively. Focusing attention on
the center plot, a more detailed picture emerges wherein the ensemble
is well-dispersed or even over-dispersed in certain locations (darker
shades of red) and under-dispersed in other regions (lighter shades of
red to white). As such, we were unable to use this diagnostic to
improve the performance of the probabilistic ML methodology to the
extent we originally anticipated. Nevertheless, we find this is a
useful diagnostic that succintly characterizes the behavior of the
ensemble predictions produced by a probabilistic ML framework.

\begin{figure}[H]
    \centering
    \includegraphics[width=0.95\textwidth]{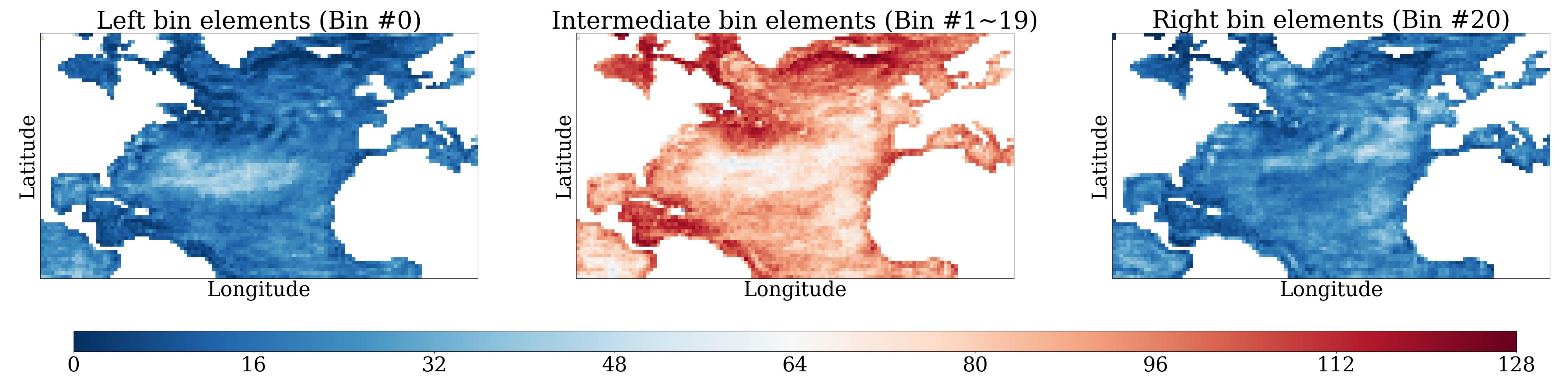}
    \caption{Heatmap of particle number per location.}
    \label{fig:heatmap}
\end{figure}

\section{Conclusions}
\label{sec5}
Following concerted national and international efforts over the past
70 years to model climate, comprehensive climate models have emerged
as a powerful tool in helping unravel and better understand the myriad
processes underlying climate and climate change. For example, such
models now help us better anticipate the climate system’s response to
{\em external} forcings, such as those due to increased greenhouse
gases on timescales longer than a few decades. However, efforts aimed
at nearer term predictions are still only in a nascent stage given the difficult
the comprehensive climate models have in representing and capturing
{\em internal} modes of variability that are relevant at these shorter
timescales with adequate accuracy. Furthermore, comprehensive
climate models demand extensive infrastructure and are computationally
very intensive. In this context, the increasing reliance on
predictions of future climate for a wide variety of purposes ranging
from integrated assessment to developing mitigation strategies to
developing resilience and adaptation strategies, makes the
availability of computationally efficient and accurate surrogates of
comprehensive earth system models highly attractive.

In this paper we add to the growing body of efforts to build
surrogates by first considering a recently proposed convolutional
network architecture to develop such a surrogate and then integrating
Bayesian inference into this architecture to further assess predictive
uncertainty. We show that the resulting Bayesian deep learning model
while marginally improving prediction accuracy, also provides a
quantification of the uncertainty inherent in the data and that
arising from the model itself, on having considered a particular
architecture (inductive bias).

The probabilistic climate prediction framework we develop has multiple
uses including obtaining information of alternative future evolutions
and the possibility of predicting extreme events. However, given the
experimental nature of the work, we went on to use diagnostics
developed in the context of probabilistic weather prediction to
examine the quality of the probabilistic ML predictions and its
utility in providing insights into the methodology itself. The use of
such diagnostics allowed us to examine certain characteristics of the
prediction ensemble such as its reliability---a property that permits
the use of the ensemble spread to estimate prediction error. Indeed,
we find that the error-spread relation of the prediction ensemble is
not optimal and suggest that efforts to drive such diagnostic
relationships to optimality is one way to improve the probabilistic ML
methodology itself.

\section*{Acknowledgments}
\label{sec6}

%Text here ===>>>
BN was supported by the U.S. Department of Energy (DOE), Office of
Science's Scientific Discovery through Advanced Computation (SciDAC)
program under project ". All other authors were supported by the US DOE, SC's
Office of Advanced Scientific Computing Research under Award Number
DE-SC-0012704. Brookhaven National Laboratory is supported by the
DOE’s Office of Science under Contract No. DE-SC0012704. This research used Perlmutter supercomputer of the National Energy Research Scientific Computing Center, a DOE Office of Science User Facility supported by the Office of Science of the U.S. Department of Energy under Contract No. DE-AC02-05CH11231 using NERSC award NERSC DDR-ERCAP0022110.

\appendix
\section{Network Architecture}
\label{app1}
This appendix discusses details related to the network architecture used in the represented case study. After an extensive hyperparameter search, Table A1 reflects the most promising fully connected convolutional neural networks configuration. As discussed in Section \ref{sec32}, we added fully connected linear layers at the bottleneck, and the modified network follows the structures specified in Table \ref{DL_beta}. In both tables, $k$ denotes the size of the convolving kernel, $s$ represents the stride of the convolution, $p$ is the zero-padding added to both sides, $K$ indicates the growth rate in dense block, and $L$ notes the number of layers.

\begin{table}[H]
\caption{Network architecture of \texttt{DL}}
\label{DL_alpha}
\begin{center}
\begin{tabular}{lll}
\multicolumn{1}{c}{\bf Name}  & \multicolumn{1}{c}{\bf Resolution} &  \multicolumn{1}{c}{\bf Configuration}
 \\ \hline \\
Input         & $ 36  \times 70 \times 125 $  & NA    \\
Convolution   & $ 128 \times 35 \times 63  $  & $ k7s2p3 $    \\
Dense Block   & $ 176 \times 35 \times 63  $  & $ K16L3 $    \\
Downsampling  & $ 88  \times 18 \times 32  $  & $ k1s1p0 \,\, \& \,\, k3s2p1 $    \\
Dense Block   & $ 184 \times 18 \times 32  $  & $ K16L6 $    \\
Upsampling    & $ 92  \times 36 \times 64  $  & $ nearest \,\, \& \,\, k3s1p1 $    \\
Dense Block   & $ 140 \times 36 \times 64  $  & $ K16L3 $    \\
Upsampling    & $ 35 \times 70 \times 125 $   & $ nearest \,\, \& \,\, k3s1p1 $    \\
Output        & $ 1   \times 70 \times 125 $  & NA    \\
\end{tabular}
\end{center}
\end{table}

\begin{table}[H]
\caption{\texttt{DL} with MLP at the bottleneck}
\label{DL_beta}
\begin{center}
\begin{tabular}{lll}
\multicolumn{1}{c}{\bf Name}  & \multicolumn{1}{c}{\bf Resolution} &  \multicolumn{1}{c}{\bf Configuration}
 \\ \hline \\
Input         & $ 36  \times 70 \times 125 $  & NA    \\
Convolution   & $ 128 \times 35 \times 63  $  & $ k7s2p3 $    \\
Dense Block   & $ 176 \times 35 \times 63  $  & $ K16L3 $    \\
Downsampling  & $ 88  \times 18 \times 32  $  & $ k1s1p0 \,\, \& \,\, k3s2p1 $    \\
Convolution   & $ 1 \times 18 \times 32  $  & $ k3s1p1 $    \\
Linear        & $ 576  $  & NA    \\ 
Convolution   & $ 48 \times 18 \times 32  $  & $ k3s1p1 $    \\
Dense Block   & $ 96 \times 18 \times 32  $  & $ K16L3 $    \\
Concatenation & $ 184 \times 18 \times 32  $  & NA    \\
Upsampling    & $ 92  \times 36 \times 64  $  & $ nearest \,\, \& \,\, k3s1p1 $    \\
Dense Block   & $ 140 \times 36 \times 64  $  & $ K16L3 $    \\
Upsampling    & $ 35 \times 70 \times 125 $   & $ nearest \,\, \& \,\, k3s1p1 $    \\
Output        & $ 1   \times 70 \times 125 $  & NA    \\
\end{tabular}
\end{center}
\end{table}

% \section{Temporal Correlation}
% \label{app2}

% \begin{figure}[H]
%     \centering
%     \includegraphics[width=0.95\textwidth]{figures/cor.pdf}
%     \caption{(a) Interval of autocorrelation of all ocean grids; and (b) The selected high temporal correlated area.}
%     \label{fig:app1}
% \end{figure}

\bibliography{ref}

\end{document}